\documentclass[]{article}

\usepackage{color}
\usepackage[usenames,dvipsnames,svgnames,table]{xcolor}
\usepackage{graphicx}

\small

\begin{document}

\twocolumn

\begin{center}
\fboxrule0.02cm
\fboxsep0.4cm
\fcolorbox{Brown}{Ivory}{\rule[-0.9cm]{0.0cm}{1.8cm}{\parbox{7.8cm}
{ \begin{center}
{\Large\em Perspective}

\vspace{0.5cm}

{\Large\bf EXor phenomenon}

\vspace{0.2cm}

{\large\em Dario Lorenzetti}


\vspace{0.5cm}

\centering
\includegraphics[width=0.25\textwidth]{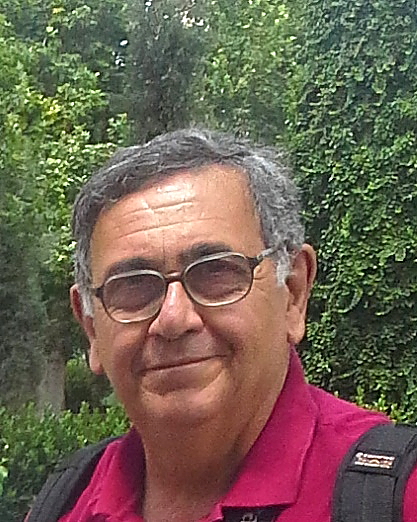}
\end{center}
}}}
\end{center}

\normalsize

{\Large\bf 1. Introduction}

Young stellar objects (YSOs) of low-to-intermediate mass (0.5-8 M$_\odot$) accumulate most of their final mass during the so-called main accretion phase, lasting about 10$^5$yrs. After this stage, the stars appear as a pre-main sequence objects and the mass accretion process continues at lower rates from their circumstellar disks. According to a widely accepted picture, the material moves through the viscous disk and eventually falls onto the star surface following the magnetic interconnection lines (Shu et al. 1994).
Observations show that the disk accretion process takes place through rapid and intermittent outbursts, usually detected at optical and near-IR wavelengths, which can be related to a sudden increase of the mass accretion rate by orders of magnitude (e.g. Hartmann \& Kenyon 1985). Although only a minor fraction of the total mass is accumulated during these last disk accretion events, nevertheless an in-depth study of them is crucial to understand how the accretion process eventually halts (thus determining the observed Initial Mass Function), and also because these bursts substantially alter the physical and chemical properties of the circumstellar environment, which has major effects on the formation of proto-planetary systems around young stars.\

A small and irregular photometric variability (typically 0.2-1 mag) attributable to disk accretion is a defining feature of all the classical T Tauri stars (CTTSs). However, several young sources display powerful outbursts of much larger intensity (up to 4-5 mag).
These objects are usually classified into two major classes (see also the review by Audard et al. 2014): 
(1) FUors (Hartmann \& Kenyon 1985) characterized by bursts of long duration (tens of years) with accretion rates of the order of 10$^{-4}$-10$^{-5}$ M$_{\odot}$~yr$^{-1}$ and spectra dominated by absorption lines; 
(2) EXors (Herbig 1989) with shorter outbursts (months--one year) with a recurrence time of years, showing accretion rates of the order of 10$^{-6}$-10$^{-7}$ M$_{\odot}$~yr$^{-1}$, and characterized by emission line spectra (e.g. Herbig 2008, Lorenzetti et al. 2009, K\'{o}sp\'{a}l et al. 2011, Sicilia-Aguilar et al. 2012, Antoniucci et al. 2013a). Such features make EXors the ideal candidates to investigate their quiescence vs. outburst properties with the same instrumentation on relatively short time scales, thus allowing  evolutionary studies (based on individual objects) more than statistical ones (based on a class of different sources).\

So far, around two dozens of EXor systems are known which can be grouped into two sub-classes: the classical EXor
(Herbig 1989, 2008), and the new identifications (see Audard et al. 2014 and Lorenzetti et al. 2012 for the list of references).  Historically, the first group was identified in the visual band and, consequently, these accretion-disk systems
are essentially unobscured (i.e. without significant envelopes). Later, the increasing
availability of near-IR facilities quite naturally favoured the identification of
more embedded eruptive variables, typically associated with an optical-IR nebula (second sub-class).
The membership to this latter class often relies on sporadic events and does not stem from a comprehensive analysis aimed at
checking whether the object presents all the properties typical of the classical prototypes (i.e. repetitive outbursts, rapid brightening and slower fading, colors pre-and post-outburst). In this scenario, the classical EXor, having already emerged
from their dust cocoons, might be associated with later phases of the pre-main sequence evolution.\

\begin{figure}[ht!]
\centering
\includegraphics[width=0.49\textwidth]{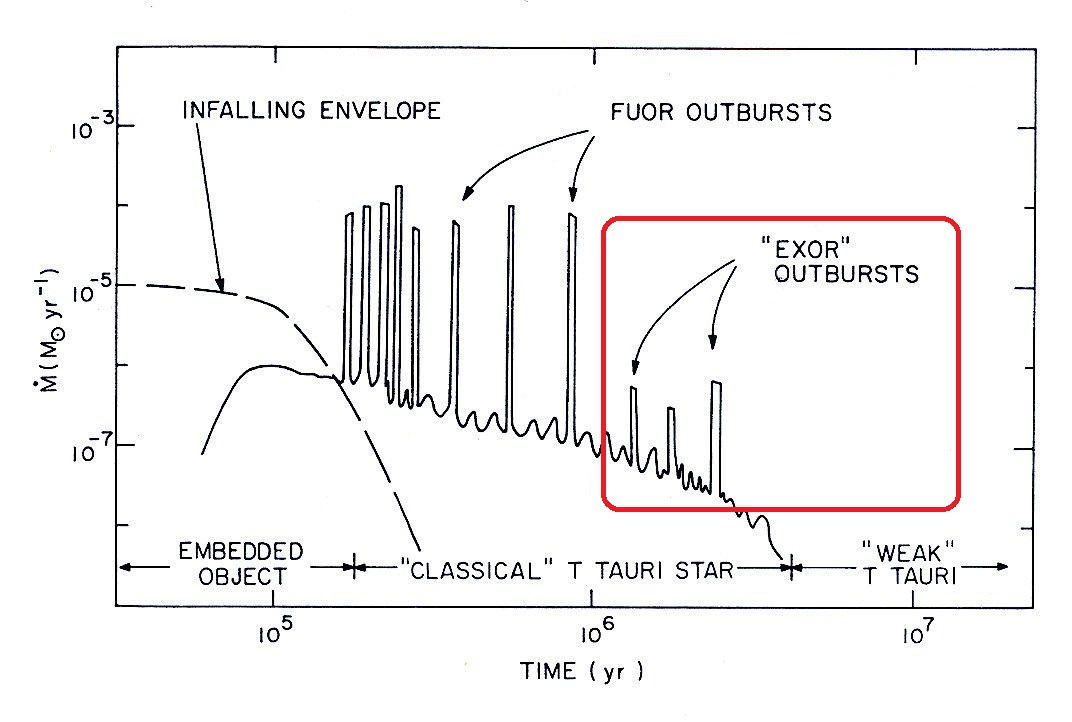}
\caption{Schematic illustration of the variation of the mass accretion rate vs. time for a young star. 
EXor events are framed in red. (\textit{Adapted from Hartmann et al. 1993}).}
\label{fig1:fig}
\end{figure}

Currently, doubts exist about the classification as EXor or FUor of some recently detected outbursts (see e.g. the case 
of V2493 Cyg - Miller et al. 2011; Semkov \& Peneva 2010; K\'{o}sp\'{a}l et al. 2011). Sometimes even the origin of
the observed variability (accretion or extinction driven) is matter of debate (see e.g. the case of GM Cep - Sicilia-Aguilar
et al. 2008; Xiao et al. 2010). These dubious circumstances arise from observations often relying on a single event and not on a long lasting photometric and/or spectroscopic monitoring. 
The nature of EXors is still very uncertain: \textit{i}) they resemble CTTSs in quiescence (e.g. Lorenzetti et al. 2007, Sipos et al. 2009); \textit{ii}) they could represent an intermediate stage between the strong FUor eruptions and the more quiescent CTTSs, or, alternately, an enhanced version of the latter class (see Figure~\ref{fig1:fig} - Herbig 1977; Hartmann et al. 1993); \textit{iii}) one could speculate that EXor events represent an infrequent manifestation of a rather common phenomenology displayed by all CTTSs. In this context, the rarity of EXor objects could be due to the lack of systematic monitoring.\

{\bf EXORCISM: A Monitoring Program}\

The very uncertain picture that emerges when interpreting the EXor events depends not only on the small number of known EXor objects, but stems also from the lack of a proper long-term multi-wavelength monitoring and 
from the fact that only a limited number of studies were able to analyse photometry and/or spectroscopy of the outburst phase and compare it (at least partially) to quiescence phases. For these reasons, our group started an observational program dubbed EXORCISM (EXOR optiCal and Infrared Systematic Monitoring - Antoniucci et al. 2014) that is intended to perform a photometric and spectroscopic monitoring in the range 0.4-2.5 $\mu$m of about 30 objects identified as known eruptive variables or candidates. Some results of this project will be commented in the following.

\vspace{0.5cm}

{\Large\bf 2. Theoretical View}\ 

Indeed, no detailed analysis or modelling of the disk structure in EXors has been performed so far, thus all the theoretical
ideas cited below originate from efforts to explain the FUor events. Nowadays the most successful model
remains that by Hartmann \& Kenyon (1985): accretion matter viscously migrates through the circumstellar disk toward its inner edge from where this matter is intermittently channelled along the magnetic field lines to the central star; the fall onto the stellar surface produces a shock that cools by emitting a hot continuum. More recently, D'Angelo \& Spruit (2010) provided quantitative predictions for the episodic accretion of piled-up material at the inner edge of the disk whenever this crosses the co-rotation radius; they also give indication that the cycle time of the bursts increases with a decreasing accretion rate. However, the mechanism responsible for the onset of EXor accretion outbursts is not known: two proposed scenarios involve essentially \textit{i}) disk instability and \textit{ii}) perturbation (of the disk) by an external body. While gravitational instabilities (e.g. Adams \& Lin 1993) is unlikely to account for the observed short time-scale of the photometric variations, the thermal ones (Bell $\&$ Lin 1994) offer a more acceptable explanation. They occur when the disk temperature reaches a value of about 5000 K and its opacity becomes a strong function of any (even very small) temperature fluctuation. Thermal models, however, present some difficulties and limitations mainly dealing with the unrealistic constraints to impose on the disk viscosity. As a consequence, other relevant mechanisms have been considered, such as the presence of a massive planet that opens up a gap in the disk (Lodato $\&$ Clarke 2004) able to trigger thermal instabilities at the gap itself. Alternatively, accretion bursts can be caused by an external trigger such as a close encounter in a binary system (Bonnell \& Bastien 1992; Reipurth \& Aspin 2004). Indeed, many EXor are binary systems and a small body has been recently detected by K\'{o}sp\'{a}l et al. (2014) as close as 0.06 AU from the central star EX Lup that is considered just the prototype of the EXor class.

\vspace{0.5cm}

{\Large\bf 3. Observational View: quiescence vs. outburst}\ 

Large part of the literature on EXor systems concerns observational studies typically dealing with individual objects, whereas fewer papers investigate significant (sub-)sam\-ples of the whole class; nevertheless it is still possible to recognize some common traits among different sources. In the following a limited set of observational data will be mentioned by assembling them per spectral band and
trying to identify the relevance of each of them in building-up a consistent framework of the EXor phenomenology.\\

{\bf 3.1 High-Energy}\ 

Although strong X-rays emission is typical for young stars (Feigelson \& Montmerle 1999), few data exist for the EXor variables. 
In particular, debate exists on whether the X-rays emission is due to the accretion shock and, as such, correlates with
outburst and fading, or, alternatively, it is a result of the coronal activity and thus unrelated to any accretion-driven fluctuations. Unambiguous results have been obtained by Kastner et al. (2004) and Teets at al. (2012), who monitored the source V1647 Ori during the increasing and decreasing phases, respectively. A strong correlation is found between the decreasing optical and X-ray fluxes following the peak of the outburst in the optical, which suggests that the declining fluxes in both bands are the result of a declining accretion rate. However, the membership of V1647 Ori to the EXor class (instead of the FUor one) is debated (Aspin et al. 2011). The X-rays observational scenario for sources recognized as true EXors is not any clearer. The prototype EX Lup shows a soft X-rays spectral component most likely associated with accretion shocks. The hard X-rays component is most likely associated with a smothered stellar corona (Grosso et al. 2010). The return to quiescence of the classical EXor V1118 Ori displayed a correlation between the decreasing optical/near-IR and X-rays fluxes (Lorenzetti et al. 2006; Audard et al. 2005). However, the (coronal) plasma temperature was variable with some indication of a cooling in the early phase of the outburst with a gradual return to normal values (Audard et al. 2010).
Further X-rays observations triggered by optical burst events, and thus simultaneous (as much as possible) with them, are in order to firmly ascertain whether or not plasma temperature can be considered a parameter sensitive to the mass accretion.\

{\bf 3.2 Optical and near-IR}\

\begin{figure}[ht!]
\centering
\includegraphics[width=0.50\textwidth]{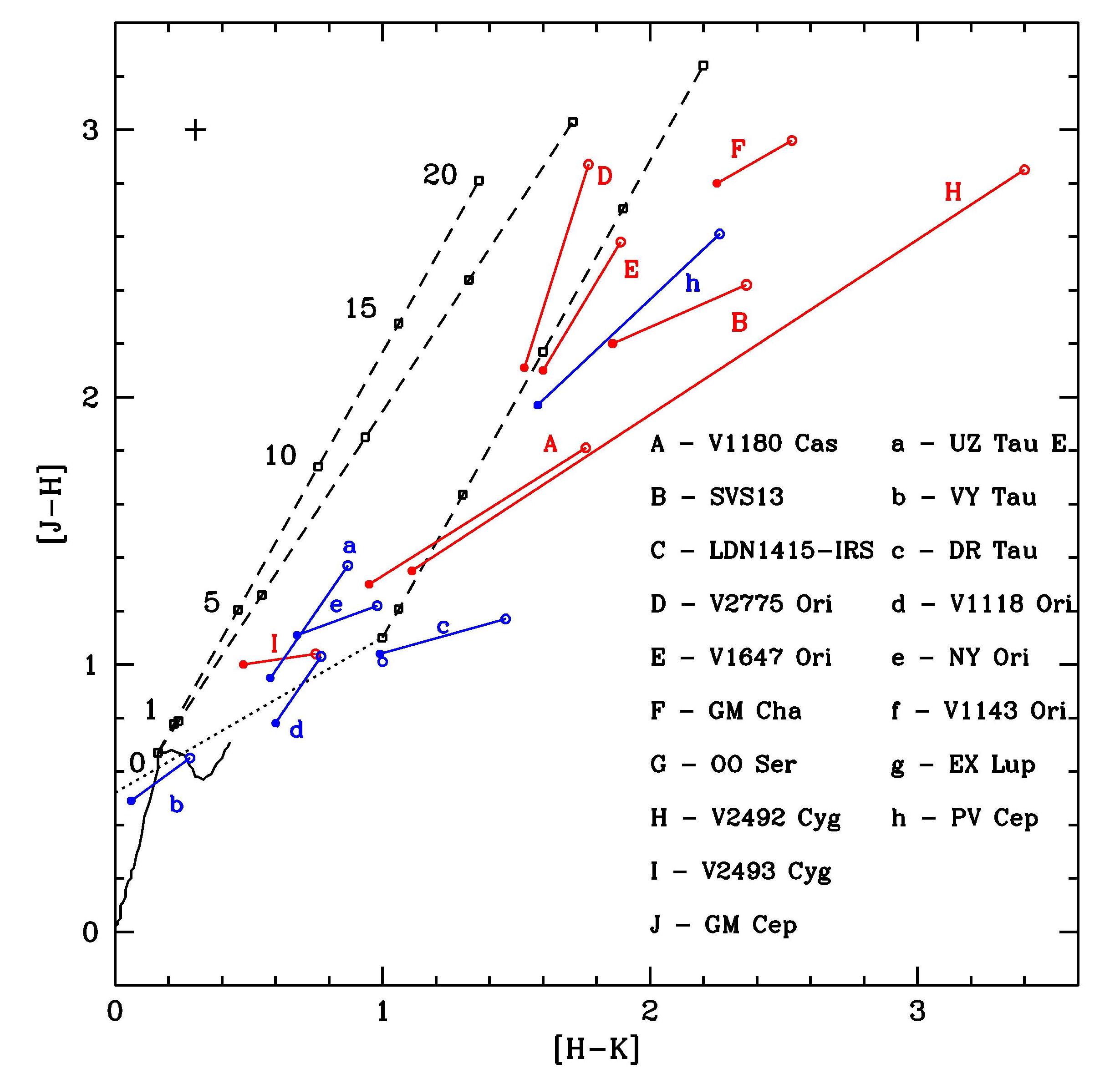}
\caption{Near-IR two-colors plot of classical (blue) and new candidates (red) EXor in two different epochs.
The solid black line (low left corner) marks the unreddened main sequence, whereas the dotted one is the locus of the T Tauri stars (Meyer et al. 1997). The dashed lines represent two reddening laws (Rieke \& Lebofsky 1985 and Cardelli et al. 1989). Solid (open) circles identify the outburst (quiescence) colors. Cassical EXor overlap much better with the T Tauri locus.}
\label{fig2:fig}
\end{figure}

{\bf \textit{3.2.1 Photometry}} - Optical (UBVRI) and near-IR (JHK) photometric data are widely used to build-up:\\ 
{\bf - light-curves of individual EXors}. Such plots illustrate the different phases of the outburst process: \textit{i}) duration and recurrence of the events (e.g. Aspin et al. 2010, 2011; Hillenbrand et al. 2013), sometime evidencing how the flux increase is typically more rapid then the subsequent fading (e.g. K\'{o}sp\'{a}l et al 2011); \textit{ii}) the appearance of spikes, attributed to hot/cool spots onto the stellar surface (Tackett et al. 2003);  \textit{iii}) the shapes of the flux variations that allow to distinguish between true EXors and UXors  objects (e.g. Grinin et al. 2000; K\'{o}sp\'{a}l et al. 2013; Miller et al. 2011); \textit{iv}) clues of some temporal lag shown by two EXors  between light-curves in different bands (Lorenzetti et al. 2011), that, if confirmed by future observations, indicates that a common property (e.g. grains thermal capacitance) regulates the emitting matter response of all EXor sources; \textit{v}) a small scale variability related to some periodicity, intermittent obscuration by geometrical effects are often observed in connection with accretion events, but it will be not discussed in the following.\\ 
{\bf - Spectral Energy Distributions (SEDs)}. Given the intrinsic nature of an EXor system (typically composed of a visible star and its circumstellar accretion disk), optical and near-IR photometry form an important part of their SEDs.
Very often, quiescence and outburst SED are jointly presented to evaluate the differences between both states: L$_{bol}$ typically increases from 1-2 (or even fractional values) to tens of L$_{\odot}$ (e.g. the cases of V2493 Cyg - K\'{o}sp\'{a}l et al. 2011,  V1118 Ori - Lorenzetti et al. 2006, EX Lup - Sipos et al. 2009). Also the shape of the SED changes, becoming bluer while the source brightens. The differential SED between the outburst and the quiescence can be well fitted with a single blackbody component with temperatures varying from 1000 K to 4500 K and emitting radii of 0.01 to 0.1 AU, as if an additional thermal component appears during the outburst phase (Lorenzetti et al. 2012). A hotspot model can fit the outburst SED, although large coverage factors are required (Audard et al. 2010). To extend such analysis to the entire sample is mandatory to better investigate the origin of the observed evidence. Notably, many EXors are optically invisible in quiescence, suggesting that a variable extinction may have also a role.\\
{\bf - color-color and color-magnitude diagrams} are reliable and widely used tools to evaluate the main mechanism(s) responsible for the SED evolution. Optical and near-IR plots both show that all the EXors (classical and candidates) are bluer when brighter, likely indicating that a hotter stellar component prevails during the most active phases while a colder disk dominates the least active phases. However, the plots based on optical colors are sensitive to the extinction parameter A$_V$ toward the objects, hence well suited to determine its variations during the different phases. Conversely, near-IR diagrams ([J-H] vs. [H-K] - see Figure~\ref{fig2:fig}) clearly demonstrates that those color variations are not related to extinction phenomena, but are likely due to temperature stratifications (Lorenzetti et al. 2012). Two main differences are recognizable between classical and new EXors: these latter occupy a redder portion of the plot, and their color fluctuations are larger. The first is an obvious consequence of their location within more obscured regions, being often associated with IR nebulae (Connelley et al. 2007); the second one stems from the fact that the new source events were followed entirely in the near-IR from quiescence to outburst, thus providing the full (i.e. maximum) range of color variations.

\vspace{0.2cm}

{\bf \textit{3.2.2 Spectroscopy}} - A large database of low ($\mathcal{R}$ $\stackrel {<}{_{\sim}}$ 2000) and high ($\mathcal{R}$ $\stackrel {>}{_{\sim}}$ 10000) resolution spectra (both optical and near-IR) has been collected so far that allows a meaningful comparison between pre-outburst, outburst, and post-outburst phases.\\ 
{\bf - Optical spectroscopy} - Collection of many spectra of the same source (see e.g. the cases of the prototype EX Lup - Sicilia-Aguilar et al. 2012 and the new candidate V2492 Cyg - Hillenbrand et al. 2013). In both quiescence and outburst many permitted emission lines are present 
(e.g. CaII, HeI), which (together with Balmer HI recombination lines) are used as a proxy for deriving estimates of the accretion luminosity (L$_{acc}$) and the mass accretion rate (Alcal\'{a} et al 2014; Antoniucci et al. 2011): typical values of this latter
parameter span from 10$^{-10}$-10$^{-8}$ M$_{\odot}$~yr$^{-1}$ in quiescence to 10$^{-8}$-10$^{-6}$ M$_{\odot}$~yr$^{-1}$ in outburst (Audard et al. 2010; Lorenzetti et al. 2015; Sicilia-Aguilar et al. 2012). An active chromosphere can explain the metallic neutral and ionized lines (mostly FeI and FeII) typical of quiescence and outburst narrow component (v$_{peak}$ = 5-10 km/s), whereas broad components profiles 
(v$_{peak}$ = 50-200 km/s) suggest they originate in hot dense accretion column that suffers velocity variations. The similarity between the pre-outburst and post-outburst states suggest that the accretion channels are similar during the whole period, and only the accretion rate varies. Photospheric signatures (typically LiI $\lambda$6707) observed at high S/N are typically highly veiled, hence used to evaluate this parameter that shows values up to 20 (Alencar et al. 2001). The most embedded EXors display 
several TiO/VO band heads in emission, indicating an amount of dense ($>$ 10$^{10}$ cm$^{-3}$), warm (1500-4000 K) circumstellar material (Covey et al. 2011).

\begin{figure}[ht!]
\centering
\includegraphics[width=0.50\textwidth]{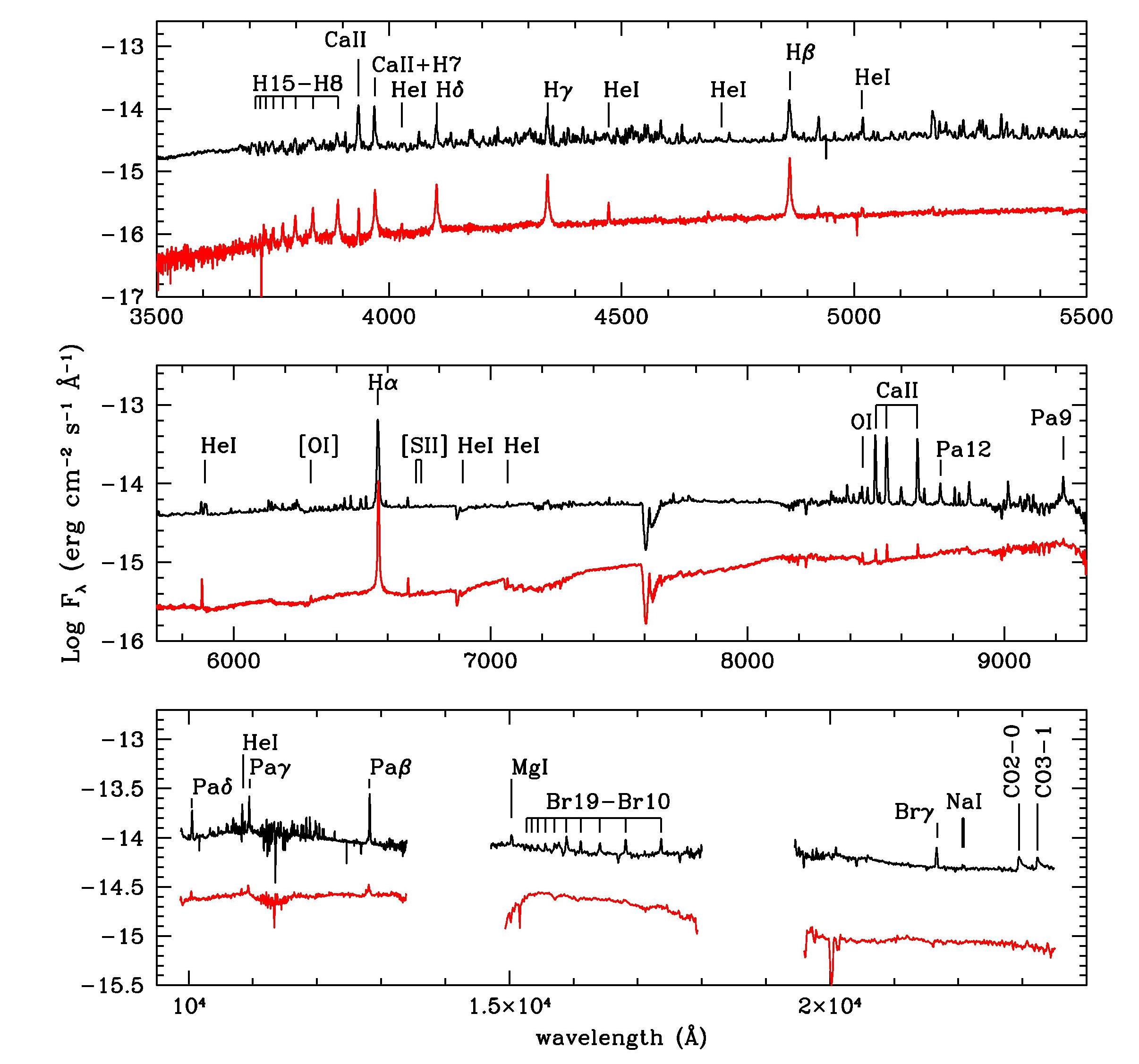}
\caption{Optical (LBT/MODS) and near-IR spectrum (LBT/LUCIFER) of V1118 Ori in outburst (black) shown in comparison with the quiescence spectrum (red - Lorenzetti et al. 2015). (\textit{from Giannini et al. 2016}).}
\label{fig3:fig}
\end{figure}

During the outburst, the number of emission lines increases dramatically, showing strong intensity variations and largely asymmetric profiles that testify a dynamic interaction between star and disk (Aspin \& Reipurth 2009). P-Cyg profiles (both direct and inverse) indicate the presence of ejection and accretion flows at velocities up to hundreds km/s. Accurate correlations between accretion and ejection signatures investigated at high spectral resolution will be fundamental
to ascertain quantitative relationships between both phenomena. Forbidden line emission (such as [OI], [SII], [FeII]) is often detected on the continuum source position and in some cases also offset from that. These lines exhibit blueshifted profiles and trace the shocked gas accelerated by protostellar winds and jets which are usual by-products of the accretion process.\\ 
{\bf - Near-IR spectroscopy}\\
In this spectral range one finds the Paschen and Brackett HI recombination lines that represent a complementary probe to the Balmer ones to investigate the mass accretion rate variability, especially for highly obscured EXors. In such cases (see e.g. V1647 Ori - Aspin et al. 2008), near-IR spectroscopy is the only mean to identify photospheric absorption features that allow to constrain the classification of the young star. Pa$\delta$ and Br$\gamma$ emission lines share the same upper levels (7-3 and 7-4 transitions, respectively), hence, neglecting collisional effects, their strength ratio depends only on atomic physics parameters. This is the reason why such ratio provides an estimate of the extinction (A$_V$) toward the star (Covey et al. 2011). The same is true for the [FeII] lines at 1.25 and 1.64 $\mu$m, but their forbidden nature make them ideally suited to trace the extinction in less dense
and shocked regions (jets, HH objects) often found around EXors (Aspin et al. 2008, Reipurth \& Bally 2001, Giannini et al. 2015).
HeI 1.08 $\mu$m provides evidence for a substantial wind associated with the outburst of both classical (DR Tau - Edwards et al. 2003) and new EXors (V2492 Cyg - Covey et al. 2011), showing strong blueshifted absorption out to about 300 km/s which dips to more than 50\% of the continuum level. Concerning the molecular lines, H$_2$ 1-0 S(1) quadrupole transition at 2.12 $\mu$m and CO overtone bandheads 2-0 and 3-1 at 2.29 and 2.32  $\mu$m, respectively, are often recurring features. The first line (in emission) traces possible HH objects close to the star and emitted as a consequence of an accretion event. CO is present in EXor spectra both in absorption and in emission: it is believed to originate in the gaseous inner disk where it traces zones relatively warm (being CO completely dissociated at $\sim$4000 K) at high densities ($>$10$^7$ cm$^{-3}$). As the outburst proceeds toward the quiescence, the spectrum displays strong CO emission, absence, and finally CO absorption. Further observations should clarify whether we are observing the internal heating of the disk that dominates the radiation transfer, or simply different regions (emission in the inner disk and absorption in the stellar photosphere (Aspin et al. 2008; Hillenbrand et al. 2013; Biscaya et al. 1997). Figure~\ref{fig3:fig} depicts optical and near-IR spectra of the representative EXor V1118 Ori obtained during its very recent 2015 outburst (Giannini et al. 2016). The comparison with the quiescence spectra (obtained with very similar instrumentation) illustrates the variations mentioned above.

{\bf 3.3 Mid- and Far-IR}\\

\begin{figure}[ht!]
\centering
\includegraphics[width=0.40\textwidth]{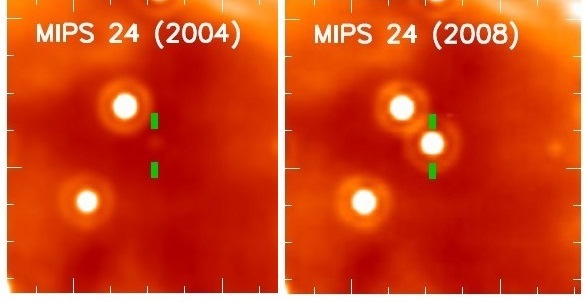}
\caption{MIPS 24$\mu$m image of the  field in two different epochs. (\textit{from Safron et al. 2015}).}
\label{fig4:fig}
\end{figure}

High sensitivity observations of EXors in this spectral region are not as much as common at shorter wavelengths, since the limited lifetime of IR satellites (ISO, $Spitzer$, WISE, Herschel) combined with the cadence and duration of the EXor events, did not allow to follow their entire evolution. Nevertheless, one specific source (V2492 Cyg) has been studied in detail (Aspin 2011; Covey et al. 2011; Hillenbrand et al. 2013; K\'{o}sp\'{a}l et al. 2013). Enlarging its available SED (up to 100-200 $\mu$m) more accurate parameters are derived by model fitting and the mid- to far-IR variability demonstrates that it is not exclusively related to changing accretion, but also to asymmetric structures in the inner disk: V2492 Cyg exhibits both accretion- and extinction-driven high-amplitude variability phenomena (more reminiscent of UXor objects). Beside this specific case, surveys of more numerous samples have been presented. The results of an $ISOPHOT$ survey of young sources (including 7 EXors) carried out by K\'{o}sp\'{a}l et al. (2012) suggest that mid-IR variability is more ubiquitous than was known before. Interpreting this variability is a new possibility for exploring the structure of the disk and its dynamical processes. A complete $WISE$ (3.4-22 $\mu$m) survey of all the EXor has been presented (Antoniucci et al. 2013b) showing SED's compatible with the existence of an inner hole in the circumstellar disk. This compilation is intended as a first step toward the construction of a significant database in this spectral regime. 
As noted above, any discovery of a new EXor outburst has been done so far in the optical/near-IR band although, in principle, there is no physical reason preventing the EXor phenomenum from occurring also during the more embedded (i.e. earlier) phase. Indeed,
a new outbursting object (HOPS 383), invisible at shorter wavelengths, has been recently found in the 24$\mu$m images of Orion (see Figure~\ref{fig4:fig} - Safron et al. 2015). Such a discovery, suggesting that accretion outbursts occur even in the earliest phases of the pre-main sequence evolution, could open a new investigative approach. 

{\bf 3.4 Sub-mm regime}\\

A survey of EXor at (sub-)mm wavelengths does not exist at the moment. A few objects have been individually observed: V2492 Cyg at 1.1 mm (Aspin 2011) and 2.7 mm (Hillenbrand et al. 2012); GM Cep at 1.3 mm (Sicilia-Aguilar et al. 2008); OO Ser at 800 $\mu$m (Hodapp et al. 1996). Submillimeter Array (SMA) 1.3 mm high angular resolution observations of four EXors (VY Tau, V1118 Ori, V1143 Ori and NY Ori)
(Liu et al. 2016) mostly show very low dust masses M$_{dust}$ ($\stackrel {<}{_{\sim}}$ 10$^{-5}$-10$^{-6}$ M$_{\odot}$) in the associated circumstellar disks. These findings suggest that the gas and dust reservoirs that feed the repetitive outbursts may be limited to the small-scale innermost region of their circumstellar disks (smaller than the resolution limit). ALMA will be needed to improve our understanding of the triggering mechanisms of EXor outbursts.

\vspace{0.5cm}

{\Large\bf 4. Other perspectives}\\

Open questions and observational perspectives have been mentioned through all the sections above, here some further points that could be addressed in the next future are listed.
\begin{itemize}
\item Large-field monitoring of substantial portions of already known star forming regions are fundamental to enlarge the EXor sample.
\item Models of magnetized accretion disks have been recently proposed (Lizano et al. 2016): ALMA will provide direct measurements
of magnetic fields and their morphology at disk scales. 
\item  EXor are too faint for the current sensitivity of interferometric facilities, but an improved instrumentation
(e.g. {\it LINC-NIRVANA} at LBT) combined with deconvolution algorithms of high contrast images, already developed (La Camera et al. 2014), will be used to recover information on both disk morphology and presence of close companions. 
\item Several processes, such as crystallization (Juh\'{a}sz et al. 2012), flickering (Baek et al. 2015), extinction by dust, mass loss, are not merely concomitant with matter accretion, but intimately related to it, hence advanced studies on these subjects are essential to reach a consistent view of the EXor key parameters. 
\end{itemize}

\footnotesize

{\bf References:}\\

Adams \& Lin 1993, Protostars and Planets III, 721\\
Alcal\'{a} et al. 2014, A\&A, 561, 2\\
Alencar et al. 2001, AJ, 122, 3335\\
Antoniucci et al. 2013, ApJ 782 51\\
Antoniucci et al. 2013b, NewA, 23, 98\\
Antoniucci et al. 2014, Protostars and Planets VI (poster)\\ 
Aspin et al. 2008, AJ, 135, 423\\
Aspin et al. 2010, ApJ, 719, 50\\
Aspin 2011, AJ, 142, 135\\
Aspin \& Reipurth 2009, AJ, 138, 1137)\\
Audard et al. 2005, ApJ, 635, L81\\
Audard et al. 2010, A\&A, 511, 63\\
Audard et al. 2014, Protostars and Planets VI (review)\\
Baek et al. 2015, AJ, 149, 73\\
Bell \& Lin 1994, ApJ, 427, 987\\
Biscaya et al. 1997, ApJ, 491, 359\\
Bonnell \& Bastien 1992, ApJ 401, L31\\
Cardelli et al. 1989, ApJ, 345, 245\\
Connelley et al. 2007, AJ, 133, 1528\\
Covey et al. 2011, AJ, 141, 40\\
D'Angelo \& Spruit 2010, MNRAS 406, 1208\\
Edwards et al. 2003, ApJ. 599, L41\\
Feigelson \& Montmerle 1999, ARAA 37, 363\\
Giannini et al. 2015, ApJ, 798, 33\\
Giannini et al. 2016, ApJ, submitted\\
Grinin et al. 2000, Ast. Lett., 34, 114\\
Grosso et al. 2010, A\&A, 522, 56\\
Hartmann et al. 1993 Protostars and Planets III, p.497\\
Hartmann \& Kenyon 1985, ApJ, 299, 462\\
Herbig 1977, ApJ 217 693\\
Herbig 1989, ESO Workshop Low Mass Star Formation and Pre-Main Sequence Objects, p.233\\
Herbig 2008, AJ, 135, 637\\
Hillenbrand et al. 2013, AJ, 145, 59\\
Hodapp at al. 1996, ApJ, 468, 861\\
Juh\'{a}sz et al. 2012, ApJ, 744, 118\\
K\'{o}sp\'{a}l et al. 2011, A\&A, 527, 133\\
K\'{o}sp\'{a}l et al. 2012 , ApJS , 201, 11\\
K\'{o}sp\'{a}l et al. 2013, A\&A, 551, 62\\
K\'{o}sp\'{a}l et al. 2014, A\&A, 561, 61\\
La Camera et al. 2014, PASP, 126, 180\\
Liu et al. 2016, ApJL, in press - arxiv:1512.05902\\
Lizano et al. 2016, ApJ, in press - arxiv:1512.01159\\
Lodato \& Clarke 2004, MNRAS, 353, 841\\
Lorenzetti et al. 2006, A\&A, 453, 579\\
Lorenzetti et al. 2007, ApJ, 665, 1193\\
Lorenzetti et al. 2009, ApJ, 693, 1056\\
Lorenzetti et al. 2011, ApJ, 732, 69\\
Lorenzetti et al. 2012, ApJ, 749, 188\\
Lorenzetti et al. 2015, ApJ, 802, 24\\
Miller et al. 2011, ApJ, 730, 80\\
Myers et al. 1997, AJ, 114, 288\\
Reipurth \& Aspin 2004, ApJ, 608, L65\\
Reipurth \& Bally 2001, ARAA, 39, 403\\
Rieke \& Lebofsky 1985, ApJ, 288, 618\\
Safron et al. 2015. ApJL, 800, 5\\
Shu et al. 1994, ApJ, 429, 781\\
Semkov \& Peneva 2010, Astronomer's Telegram \#2801\\
Sicilia-Aguilar et al. 2008, ApJ, 673, 382\\
Sicilia-Aguilar et al. 2012, A\&A, 544, 93\\
Sipos et al. 2009, A\&A, 507, 881\\
Tackett et al. 2003, AJ, 126, 348\\
Teets at al. 2012,  ApJ, 760, 89\\
Xiao et al. 2010, AJ, 139, 1527\\

\normalsize

\end{document}